# On Effect of Right-Half-Plane Zero Present in Buck Converters with Input Current Source in Wireless Power Receiver Systems

Kerui Li, *Student Member, IEEE*, Siew-Chong Tan, *Senior Member, IEEE,* and Ron Shu Yuen Hui, *Fellow, IEEE*

*Abstract*-In wireless power receiver systems, the buck converter is widely used to step down the higher rectified voltage derived from the wireless receiver coil, to a lower output voltage for the immediate battery charging process. In this work, the presence and effect of the right-half-plane (RHP) zeros found in the small-signal inductor-current-to-duty-ratio and output-voltage-to-duty-ratio transfer functions of the buck converter in the wireless power receiver system on the control performance, are investigated. It is found and mathematically proved that the RHP zeros are introduced by the current source nature of the system attributed to the series-series compensation and finite DC-link capacitance. The RHP zero not only results in non-monotonic open-loop dynamic response but also complicates the design of feedback control and causes potential closed-loop instability. Theoretical and experimental results are provided to validate the presence of the RHP zeros and their effect on open-loop and closed-loop dynamic responses.

## I. INTRODUCTION

The step-down wireless power receiver, comprising a front-end diode rectifier that is followed by a step-down DC-DC converter [1]-[4], is most widely used for wireless power transfer. In such applications, the buck converter is typically employed as a step-down DC-DC converter processing the higher rectified voltage derived from the wireless receiver coil, into a lower output voltage for the immediate battery charging process [1], [5].

The charging specifications of the lithium-ion battery are setting stringent output voltage/current regulation requirements on the buck converter. For example, the lithium-ion battery requires a voltage error of less than 1% (e.g. 4.25±0.03 V) during the transition of charging modes and in steady-state constant voltage charging [6]. These requirements would have been easily satisfied if the input of the buck converter is a voltage source. With the input being a voltage source, the buck converter possesses only stable poles (and no zero) in its output-voltage-to-duty-ratio transfer function. This results in its feedback control design inheriting the merits of having a high DC gain, wide closed-loop bandwidth, and a large phase margin [7]-[9]. Subsequently, good output voltage/current regulation conforming the charging requirements can be easily attained with the buck converters [10], [11].

However, the buck converter used in the wireless power receiver system does not necessarily preserve these characteristics. Owing to the use of series-series compensation in the wireless power transfer system, an independent AC current source is introduced to the input of wireless power receiver system [12], [13]. The current source nature leads to highly load-dependent steady-state DC-link and output voltages as well as small-signal dynamic behavior. Furthermore, attributed to the finite DC-link capacitance, a non-negligible phase delay is introduced to the small-signal response characteristic of the DC-link voltage. With DC-link capacitor serving as the input to the buck converter, this phase delay is further propagated to the buck converter and thus the small-signal responses of the inductor current and output voltage are coupled to the DC-link capacitance [14]-[16]. As a consequence, a buck converter operating in the wireless power receiver system features a relatively different dynamic characteristic as compared to that of conventional voltage source buck converters. Due to the popularity of using the buck converter in wireless power receiver systems [17], [18], this aspect is worth investigating.

In this work, the presence and effect of the right-half-plane (RHP) zeros found in the small-signal inductor-current-to-duty-ratio and output-voltage-to-duty-ratio transfer functions of the buck converter in the wireless power receiver system on the control performance, are investigated. The study is carried out using the time-domain and frequency-domain models of the system. Analytical and experimental study on the RHP zero as well as the open-loop and closed-loop small-signal responses of the buck converter are performed. It must be emphasized that buck converters have traditionally been perceived as systems that are easily controllable for reasons that they are minimum-phase systems that do not possess any RHP zero, regardless of whether their input voltage source is DC or rectified time-varying AC. This is well-proven in literature [7], [9], [19]. Importantly, while a voltage source buck converter with an LC input filter [7] may inherit non-minimum-phase characteristics with RHP zero, it can be eliminated by properly introducing damping resistance. Therefore, it is out of our expectation and against conventional wisdom to discover that the current-source nature of series-series compensated wireless power transfer system can actually alter the characteristic of the buck converter in the wireless power receiver system, such that it becomes a non-minimum-phase system with RHP zeros in its small-signal transfer equations. This has never been previously reported. In addition, the finding of RHP zero explains the non-minimum-phase feature of the buck converter in the wireless receiver system, and facilitates the compensator design of the system without assuming constant input voltage [17] or using a computationally-exhaustive algorithm [18].

## II. MODEL OF THE BUCK CONVERTER IN THE WIRELESS POWER RECEIVER SYSTEM

### A. Wireless Power Receiver System

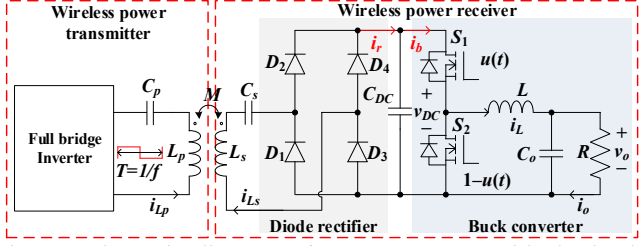

Fig. 1. Schematic diagram of a WPT system with the buck-converter wireless power receiver.

Fig. 1 shows the circuit configuration of a WPT system with the step-down buck-converter wireless receiver. The wireless power transmitter comprises a full-bridge inverter operating at switching frequency $f$ ($f=1/T$) and a primary-side transmitter coil $L_p$ with a series compensation capacitor $C_p$. The resonant frequency of the compensated coil is set equal to the switching frequency $f$, i.e., $1/\sqrt{C_pL_p} = 2\pi f$. The wireless power receiver contains the secondary-side receiver coil $L_s$ with a series compensation capacitor $C_s$ (i.e., $1/\sqrt{C_sL_s} = 2\pi f$), a first-stage diode rectifier ($D_1$ to $D_4$) for converting AC voltage into an unregulated DC voltage $v_{DC}$ at the DC-link capacitor $C_{DC}$, and a second-stage buck converter (comprising complementary switches $S_1$ and $S_2$, output inductor $L$, output capacitor $C_o$, and load $R$). The logic switching signal of switch $S_1$ is $u(t)$, and that of the complementary switch $S_2$ is $1-u(t)$. In this case,

$$u(t) = \begin{cases} 1 & \text{if } nT < t \leq (n+d)T \\ 0 & \text{if } (n+d)T < t \leq (n+1)T \end{cases} \quad (1)$$

where logic 1 and 0 respectively represent the turning ON and OFF of switch $S_1$. Here, $d$ (where $0 \leq d \leq 1$) is the duty ratio of $u(t)$ over the switching period $T$, and $n$ (where $n=0,1,..$) is an arbitrary non-negative integer. The switching frequency of $u(t)$ is synchronized at $f$ to eliminate the beat frequency oscillation [20].

### B. Operating Principle and Time-Domain Model

Considering the operation of the diode bridge and the buck converter, the wireless power receiver system has three switching states. Without loss of generality, Figs. 2 and 3 show respectively the equivalent circuit models of the wireless power receiver operating in its three switching states and the corresponding key waveforms of the system when $d \geq 0.5$. Take note that the description and subsequent analysis remains valid when $d<0.5$.

Since the series-series compensation is applied, the wireless power receiver coil $L_s$ introduces an independent sinusoidal current $i_{Ls}(t)$ to the diode bridge rectifier [12], [13], i.e.,

$$i_{Ls}(t) = I_{Ls}\sin(2\pi ft), \quad (2)$$

where $I_{Ls}$ is the amplitude of the sinusoidal input current $i_{Ls}(t)$. $I_{Ls}$ is considered as a constant value. Due to its current-source nature, the modelling and operation of the buck converter is different from those of the voltage source buck converter. The diode rectifier has two operating modes. When $i_{Ls}(t)>0$, the current flows to capacitor $C_{DC}$ via diodes $D_2$ and $D_3$. When $i_{Ls}(t)<0$, the current flows to capacitor $C_{DC}$ via diodes $D_1$ and $D_4$. Thus, the rectified current $i_r(t)$ is

$$i_r(t) = |I_{Ls}\sin(2\pi ft)| \quad (3)$$

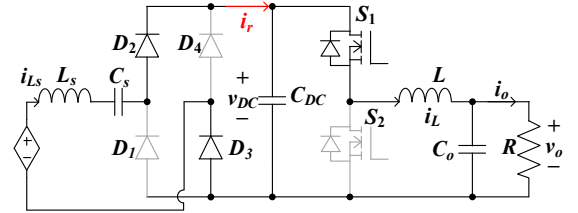

(a) Equivalent circuit model of State I

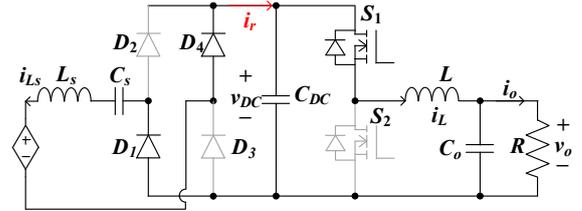

(b) Equivalent circuit model of State II

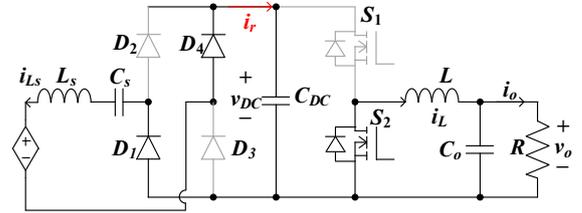

(c) Equivalent circuit model of State III

Fig. 2. Equivalent circuit models of the wireless power receiver under the three switching states.

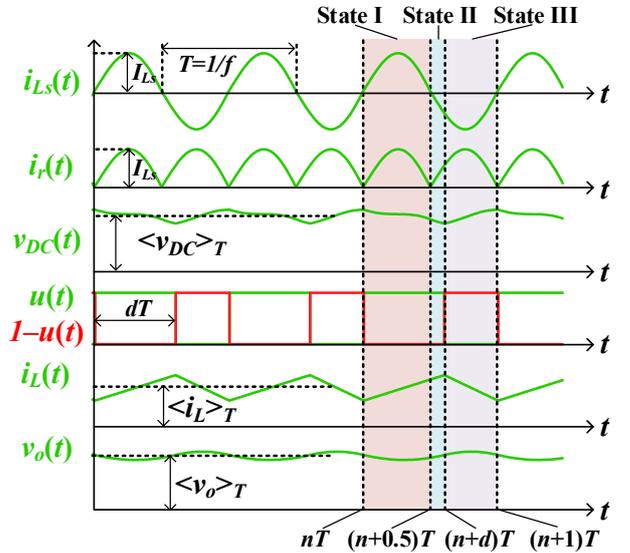

Fig. 3. Key waveforms of the wireless power receiver system.

*State* I [$nT \leq t < (n+0.5)T$]

Fig. 2(a) shows the equivalent circuit model of the power circuit in State I. During State I, $i_{Ls}(t)$ is positive and flows through $D_2$ and $D_3$. Concurrently, $S_1$ is turned ON. The differential equation of the DC-link capacitor voltage $v_{DC}(t)$ is

$$C_{DC}\frac{\partial v_{DC}(t)}{\partial t} = I_{Ls}\sin(2\pi ft) - i_L(t) \quad (4)$$

The inductor current $i_L(t)$ and output voltage $v_o(t)$ of the buck converter are governed by the differential equations

$$L\frac{\partial i_L(t)}{\partial t} = v_{DC}(t) - v_o(t) \quad (5)$$

$$C_o\frac{\partial v_o(t)}{\partial t} = i_L(t) - \frac{v_o(t)}{R} \quad (6)$$

*State II* $[(n+0.5)T \le t < (n+d)T]$:

Fig. 2(b) shows the equivalent circuit model of the power circuit in State II. During State II, $i_{Ls}(t)$ becomes negative and flows through $D_1$ and $D_4$. Meanwhile, $S_1$ remains ON. The differential equation of the DC-link capacitor voltage is

$$C_{DC}\frac{\partial v_{DC}(t)}{\partial t} = -I_{Ls}\sin(2\pi ft) - i_L(t) \quad (7)$$

The inductor current $i_L(t)$ and output voltage $v_o(t)$ are still governed by the differential equations (5) and (6).

*State III* $[(n+d)T \le t < (n+1)T]$:

Fig. 2(c) shows the equivalent circuit model of the power circuit in State III. Here, $i_{Ls}(t)$ is still negative and flows through $D_1$ and $D_4$. As $S_1$ is turned OFF, current $i_L(t)$ freewheels through $S_2$. The differential equation of $v_{DC}(t)$ is

$$C_{DC}\frac{\partial v_{DC}(t)}{\partial t} = -I_{Ls}\sin(2\pi ft) \quad (8)$$

The inductor current $i_L(t)$ and output voltage $v_o(t)$ are

$$L\frac{\partial i_L(t)}{\partial t} = -v_o(t) \quad (9)$$

$$C_o\frac{\partial v_o(t)}{\partial t} = i_L(t) - \frac{v_o(t)}{R} \quad (10)$$

### C. Averaged Model and Small-Signal Model

By applying per-switching-cycle state-space averaging [5] to (4)-(10), the averaged model is written as

$$C_{DC}\frac{\partial \langle v_{DC}\rangle_T}{\partial t} = \frac{1}{T}\int_{nT}^{(n+1)T}(i_r(t) - u(t)i_L(t))\,dt = \frac{2}{\pi}I_{Ls} - d\langle i_L\rangle_T \quad (11)$$

$$L\frac{\partial \langle i_L\rangle_T}{\partial t} = \frac{1}{T}\int_{nT}^{(n+1)T}(u(t)v_{DC}(t) - v_o(t))\,dt = d\langle v_{DC}(t)\rangle_T - \langle v_o\rangle_T \quad (12)$$

$$C_o\frac{\partial \langle v_o\rangle_T}{\partial t} = \frac{1}{T_s}\int_{nT}^{(n+1)T}(i_L(t) - v_o(t))\,dt = \langle i_L(t)\rangle_T - \frac{\langle v_o\rangle_T}{R} \quad (13)$$

where $\langle v_{DC}\rangle_T$, $\langle i_L\rangle_T$, and $\langle v_o\rangle_T$ respectively represent the average of $v_{DC}(t)$, $i_L(t)$, and $v_o(t)$ over one switching cycle $T$. Next, small-signal AC variation is introduced into the averaged variables, giving

$$\begin{cases} \langle v_{DC}\rangle_T = V_{DC} + \widetilde{v_{DC}} \\ \langle i_L\rangle_T = I_L + \tilde{i}_L \\ \langle v_o\rangle_T = V_o + \widetilde{v_o} \\ d = D + \tilde{d} \end{cases} \quad (14)$$

where $V_{DC}$, $I_L$, $V_o$, and $D$ are the respective steady-state DC values, and $\widetilde{v_{DC}}$, $\tilde{i}$, $\widetilde{v_o}$, and $\tilde{d}$ are the respective small-signal AC variations. By substituting (14) into (11)-(13) and eliminating the AC terms, the steady-state DC values can be derived as

$$\begin{cases} V_{DC} = \dfrac{2RI_{Ls}}{\pi D^2} \\ I_L = \dfrac{2I_{Ls}}{\pi D} \\ V_o = \dfrac{2RI_{Ls}}{\pi D} \end{cases} \quad (15)$$

Similarly, by cancelling the DC terms and nonlinear high-order AC terms, the small-signal models are obtained as

$$C_{DC}\frac{\partial \widetilde{v_{DC}}}{\partial t} = -D\tilde{i}_L - \tilde{d}I_L \quad (16)$$

$$L\frac{\partial \tilde{i}_L}{\partial t} = D\widetilde{v_{DC}} + \tilde{d}V_{DC} - \widetilde{v_o} \quad (17)$$

$$C_o\frac{\partial \widetilde{v_o}}{\partial t} = \tilde{i} - \frac{\widetilde{v_o}}{R} \quad (18)$$

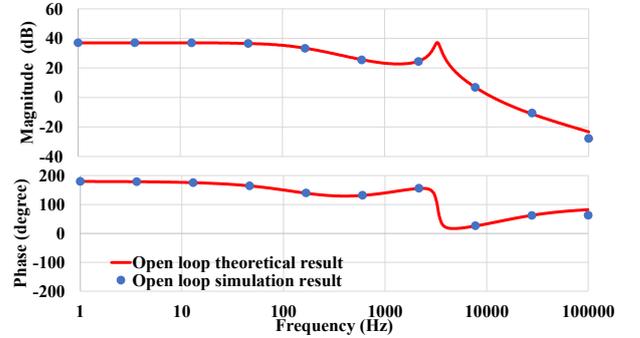

(a) Bode plots of $G_{v_{DC}}(s)$

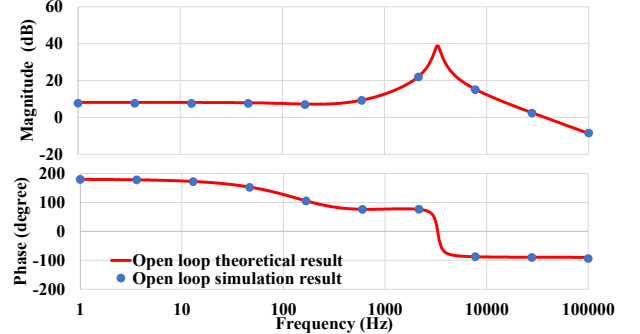

(b) Bode plots of $G_{i_L}(s)$

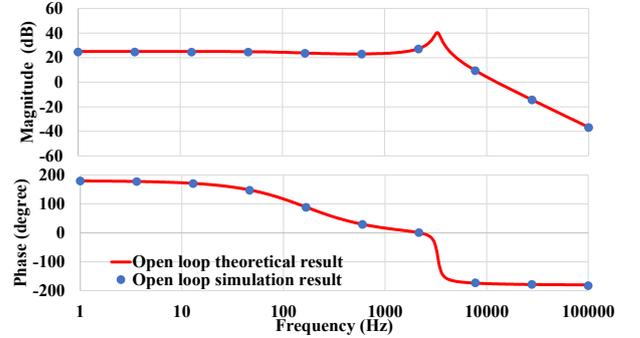

(c) Bode plots of $G_{v_o}(s)$

Fig. 4. Bode plots of the small-signal transfer functions of the buck converter in wireless receiver system.

By substituting the steady-state values (15) into the small-signal models (16)-(18), the closed-form *s*-domain small-signal transfer functions of the DC-link voltage, inductor current and output voltage with respect to the duty cycle can be derived as

$$G_{v_{DC}}(s) = \frac{\widetilde{v_{DC}}}{\tilde{d}} = \frac{-2I_{Ls}(C_o L R s^2 + (C_o R^2 + L)s + 2R)}{\pi D[C_o C_{DC} L R s^3 + C_{DC} L s^2 + (C_o R D^2 + C_{DC} R)s + D^2]} \quad (19)$$

$$G_{i_L}(s) = \frac{\tilde{i_L}}{\tilde{d}} = \frac{2I_{Ls}(C_{DC} R s - D^2)(C_o R s + 1)}{\pi D^2[C_o C_{DC} L R s^3 + C_{DC} L s^2 + (C_o R D^2 + C_{DC} R)s + D^2]} \quad (20)$$

$$G_{v_o}(s) = \frac{\widetilde{v_o}}{\tilde{d}} = \frac{2RI_{Ls}(C_{DC} R s - D^2)}{\pi D^2[C_o C_{DC} L R s^3 + C_{DC} L s^2 + (C_o R D^2 + C_{DC} R)s + D^2]} \quad (21)$$

To verify the accuracy of these models, circuit simulation of the system carried out with the PSIM V12.0.4 software and analytical results obtained from equations (19)-(21) are comparatively plotted in Fig. 4. The parameters adopted for the simulation and computation are: *f*=200 kHz, $I_{Ls}$ = 1 A, $L_p$=$L_s$=160 μH, $C_p$=$C_s$=3.96 nF, $C_{DC}$= 30 μF, *L*=77 μH, $C_o$=40 μF, *D*=0.5, $|\tilde{d}|$= 0.1, and *R*=7 Ω. As shown, the simulation and analytical results are close, thereby validating the accuracy of the small-signal model for the frequency range of 1 Hz to 100 kHz.

*D. Poles and Zeros of Transfer Functions*

Fig. 5 (a)-(c) show the pole-zero maps of the transfer functions of $G_{v_{DC}}(s)$, $G_{i_L}(s)$, and $G_{v_o}(s)$, respectively. The transfer functions share identical denominators. As a consequence, the poles of the transfer functions are identical. It is, however, difficult to obtain the closed-form expressions of the poles due to its complex third-order denominator. Nevertheless, the numerical results of the poles can be obtained via simulation as −1340+**i**20700, −1340−**i**20700, and −898.

The closed-form expression of the numerator of $G_{v_{DC}}(s)$ comprises two zeros, i.e.,

$$Z_{v_{DC},1} = \frac{-(C_o R^2 + L) + \sqrt{(C_o R^2 + L)^2 - 8C_o R^2 L}}{2C_o L R}$$

$$Z_{v_{DC},2} = \frac{-(C_o R^2 + L) - \sqrt{(C_o R^2 + L)^2 - 8C_o R^2 L}}{2C_o L R} \quad (22)$$

These zeros are located on the left-half of a complex plane, and can be computed via simulation as −7460 and −87000, respectively for $Z_{v_{DC},1}$ and $Z_{v_{DC},2}$. The left-half-plane (LHP) zeros contribute to a larger phase lag of between the input and the output of $G_{v_{DC}}(s)$.

The closed-form expression of the numerator of $G_{i_L}(s)$ also comprises two zeros, namely

$$Z_{i_L,1} = \frac{D^2}{C_{DC} R}$$

$$Z_{i_L,2} = -\frac{1}{C_o R} \quad (23)$$

Clearly, $Z_{i_L,1}$ is located on the right-half plane, whereas $Z_{i_L,2}$ is located on the left-half plane. The RHP zero $Z_{i_L,1}$ is proportional to $D^2$ and inversely proportional to $C_{DC}$ and *R*. The LHP zero $Z_{i_L,2}$ is inversely proportional to $C_o$ and *R*. The computed results of the zeros are 1190 and −3570, respectively.

The closed-form expression of the numerator of $G_{v_o}(s)$ comprise one zero

$$Z_{v_o,1} = \frac{D^2}{C_{DC} R} \quad (24)$$

Apparently, $Z_{v_o,1}$ is located on the right-half plane. It is proportional to $D^2$ and inversely proportional to $C_{DC}$ and *R*. This RHP zero limits the possible high small-signal gain at DC, wide closed-loop bandwidth and large phase margin when designing the output voltage feedback control. The computed result of the RHP zero is 1190.

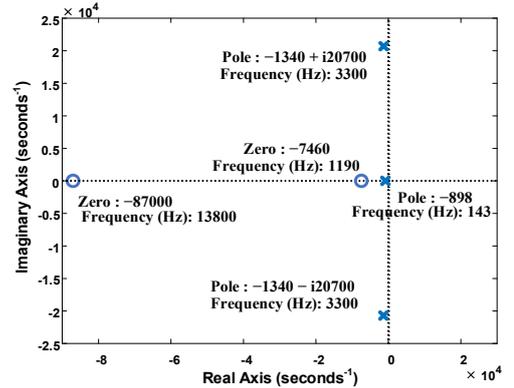

(a) Pole zero map of $G_{v_{DC}}(s)$

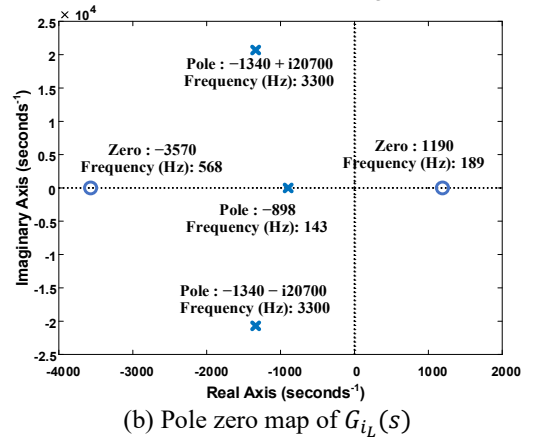

(b) Pole zero map of $G_{i_L}(s)$

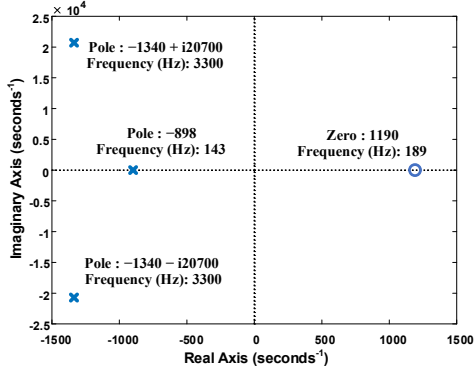

(c) Pole zero map of $G_{v_o}(s)$

Fig. 5. Poles and zeros maps of the transfer function of the buck converter.

### III. RIGHT-HALF-PLANE ZEROS OF THE BUCK CONVERTER

*A. Physical Origin and Effect of the Right-Half-Plane Zeros*

The RHP zeros presented at the transfer function of $G_{i_L}(s)$ and $G_{v_o}(s)$ are introduced by the system order transition between the three switching states shown in Fig. 2. During State I and II, switch $S_1$ is connected to the DC-link capacitor. The equivalent circuit is a third-order system. During State III, $S_1$ turns OFF and $S_2$ turns ON. The equivalent circuit is separated into one first-order system (DC-link capacitor $C_{DC}$) and one second-order system (involving inductor $L$ and output capacitor $C_o$). This system order transition, therefore, introduces the RHP zeros to $G_{i_L}(s)$ and $G_{v_o}(s)$. Moreover, the current source nature of the series-series compensated wireless power transfer system leads to a highly load-dependent steady-state DC-link and output voltages as well as small-signal dynamic behaviors. Furthermore, attributed to the finite DC-link capacitance, a non-negligible phase delay is introduced to the small-signal response characteristic of $v_{DC}$. The phase delay is further propagated to $i_L$ and $v_o$ such that the small-signal responses of $i_L$ and $v_o$ are coupled to the DC-link capacitance As a consequence, RHP zeros, are present in the model of the system.

Other than complicating feedback control design and potentially causing closed-loop instability, the RHP zeros found on $G_{i_L}(s)$ and $G_{v_o}(s)$ also result in the non-monotonic behaviour of their transient response, with correspondingly $i_L(t)$ and $v_o(t)$ having their trajectories initially heading toward the wrong direction of the control correction in the event of a duty cycle step change. This phenomenon will subsequently be explained in detail using results obtained from a circuit simulation of the condition.

Fig. 6 shows the simulated dynamic responses of $v_{DC}(t)$, $i_L(t)$ and $v_o(t)$ of the wireless power receiver (with the same simulation parameters used in Section II) where a duty ratio step change from 0.5 to 0.475 is imposed at $t=t_d$. Before the step change of duty ratio ($t<t_d$), the system is in steady state, i.e., $\partial <v_{DC}>_T/\partial t=0$, $\partial <i_L>_T/\partial t=0$ and $\partial <v_o>_T/\partial t=0$. With the reduction of duty ratio, the new steady-state values of the state variables will be higher than those of a higher $D$. That is, $V_{DC}(D=0.475)>V_{DC}(D=0.5)$, $I_L(D=0.475)>I_L(D=0.5)$, and $V_o(D=0.475)>V_o(D=0.5)$.

However, at the instance of $t=t_d^+$, when duty ratio is reduced from 0.5 to 0.475, the values of $<v_{DC}>_T$, $<i_L>_T$ and $<v_o>_T$ do not change instantly and remains constant at $V_{DC}(D=0.5)$, $I_L(D=0.5)$, and $V_o(D=0.5)$, respectively.

As a consequence, $\partial <v_{DC}(t)>_T/\partial t$ becomes positive (see (11)) and therefore $<v_{DC}>_T$ is increased. Subsequently, $<v_{DC}>_T$ keeps increasing to the new steady state with a monotonical trajectory.

In contrast, at $t=t_d^+$, the volt-second of $<i_L>_T$ attributed to the DC-link voltage is reduced. Thus, $\partial <i_L>_T/\partial t$ becomes negative (see (12)) and $<i_L>_T$ decreases. Owing to the reduction of $<i_L>_T$, $\partial <v_o>_T/\partial t$ becomes negative (see (13)) and hence $<v_o>_T$ decreases too. Gradually, as $<v_{DC}(t)>_T$ soars, $<v_{DC}(t)>_T$ contributes more volt-second to $<i_L>_T$, and $\partial <i_L>_T/\partial t$ becomes positive. Hence, $<i_L>_T$ changes direction and increases from the valley and eventually reaches the new steady state after some time. Similarly, as $<i_L>_T$ increases, $\partial <v_o>_T/\partial t$ becomes positive and $<v_o>_T$ is also increased from the valley to the new steady-state value. In essence, after the step change in duty ratio, $<v_{DC}>_T$ monotonically increases to the new steady-state value, whereas both $<i_L>_T$ and $<v_o>_T$ initially experience a drop prior to increasing to the new steady-state values with non-monotonic dynamics.

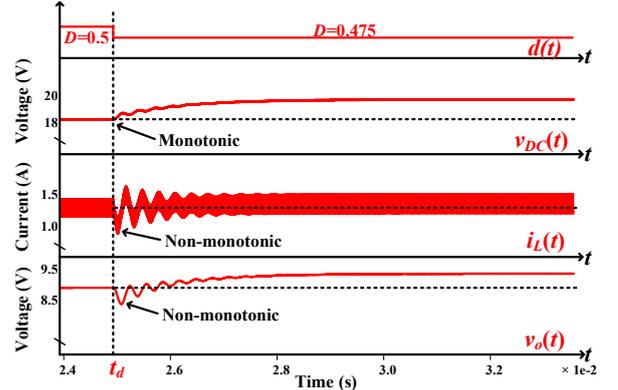

Fig. 6. Dynamic responses of of $v_{DC}(t)$, $i_L(t)$ and $v_o(t)$ after a step change on duty cycle.

*B. Analysis of Right-Half-Plane Zeros*

Fig. 7 show the pole-zero map and Bode plots of $G_{v_o}(s)$ at different operating conditions. Fig. 7(a) shows the pole-zero map and Bode plots for the condition of an increasing DC-link capacitance $C_{DC}$. In terms of pole-zero map, as $C_{DC}$ increases, both the RHP zero and the real pole approach the origin of the plane while the conjugate poles, which are at relatively high frequency, deviate away from the imaginary axis. In terms of Bode plots, as $C_{DC}$ increases, the DC gain does not change, while the phase drop due to RHP zero occurs at a lower frequency. Fig. 7(b) shows the pole-zero map and Bode plots for the condition of an increasing duty ratio $D$. In terms of pole-zero map, as $D$ increases, the real pole and RHP zero deviate away from the origin, whereas the conjugate poles move toward the imaginary axis. In terms of Bode plots, as $D$ increases, the

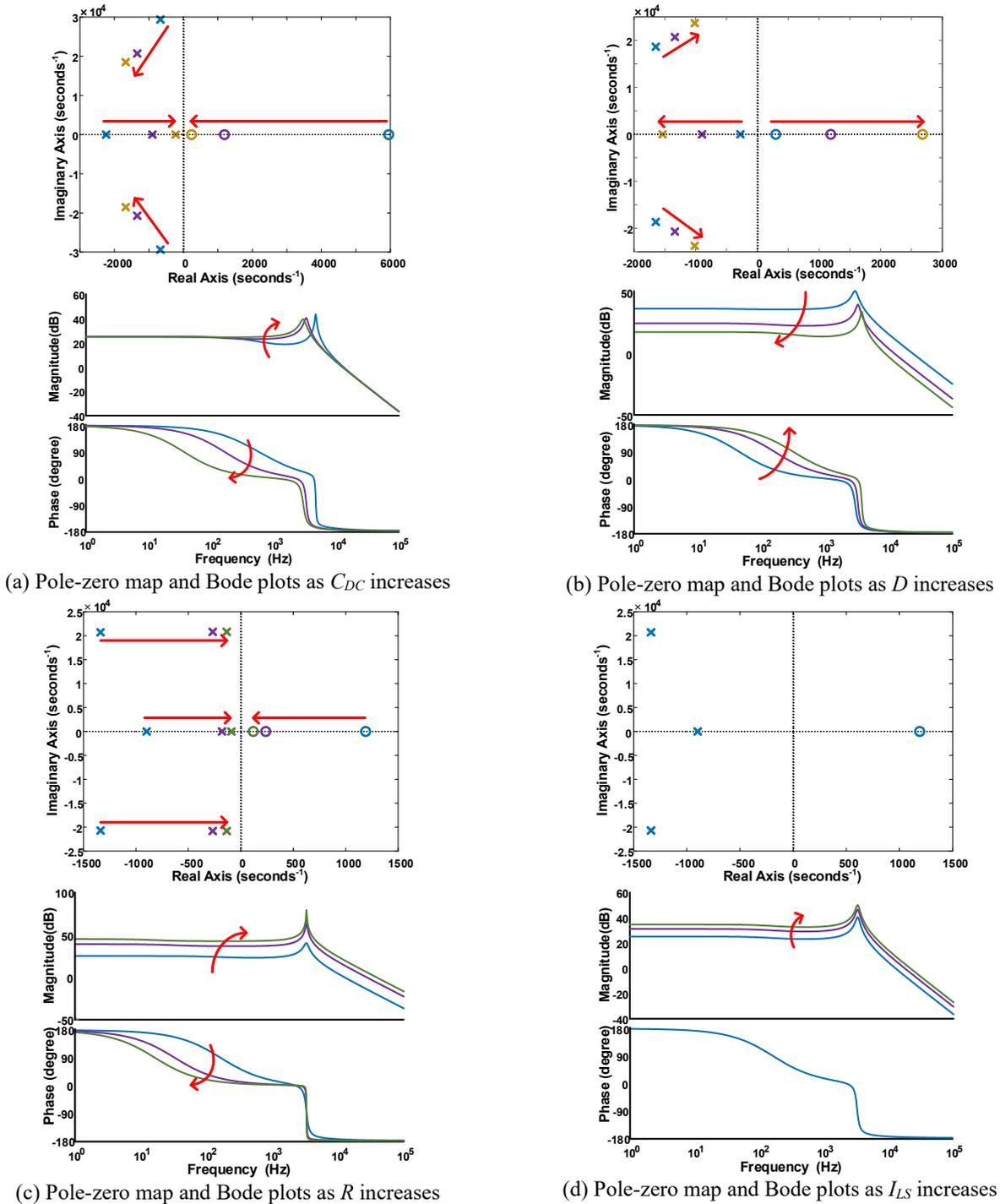

(a) Pole-zero map and Bode plots as $C_{DC}$ increases

(b) Pole-zero map and Bode plots as $D$ increases

(c) Pole-zero map and Bode plots as $R$ increases

(d) Pole-zero map and Bode plots as $I_{LS}$ increases

Fig. 7. Pole-zero map and Bode plots as $G_{v_o}(s)$ at different operating conditions.

DC gain is reduced, but the phase drop due to RHP zero shifts to a higher frequency. Fig. 7(c) shows the pole-zero map and Bode plots for the condition of an increasing load $R$. In terms of pole-zero map, as $R$ increases, the RHP zero and the real pole approaches the origin, while the conjugate poles move towards the imaginary axis. In terms of Bode plots, as $R$ increases, the DC gain is increased as well, but the phase drop due to RHP zero occurs at a lower frequency. Fig. 7(d) shows the pole-zero map and Bode plots for the condition of an increasing amplitude $I_{Ls}$ of input AC current. In terms of pole-zero map, as $I_{Ls}$ increases, the RHP zero and poles are invariant. In terms of Bode plots, as $I_{Ls}$ increases, the DC gain is increased, but the Bode phase plot remains constant. Additionally, due to the phase lagging nature of the RHP zero, the phase margin of the closed-loop system is limited. One very common solution is to push the RHP zero to the high-frequency region such that it is far beyond the bandwidth of the closed-loop system to minimize the phase lagging effect. To achieve this, one needs a small DC-link capacitor, small load resistor, or high duty ratio as illustrated in Fig. 7. Nevertheless, load resistance and duty ratio are predetermined by the required output voltage and power. Using a small DC-link capacitor is not recommended in the step-down wireless power receiver system due to beat frequency oscillations [20]. These design constraints indicate that a low-frequency RHP zero is always present in such systems. Moreover, it is worth mentioning that in addition to their existing RHP zero, extra RHP zeros will also be introduced to the buck-boost converter and boost converter of the wireless power receiver system by the input current source and the use of finite DC-link capacitor.

*C. Closed-Loop Stability*

Conventionally, the buck converter in the wireless power receiver system is regarded as an independent voltage source buck converter [17] and it is modelled as a minimum phase system (i.e., no RHP zero). However, the actual presence of the RHP zeros in the system will deteriorate the stability margin and can more easily lead to control instability if the control feedback design is not properly considered. To understand the effect of closed-loop stability under such a condition, the examination of the output voltage regulation performance of the converter with the proportional-integral (PI) compensator, is conducted. The transfer function of the PI compensator $G_c(s)$, given as shown, is

$$G_c(s) = \frac{k_p\left(s + \frac{k_i}{k_p}\right)}{s} \quad (25)$$

The PI compensator derived from the voltage source buck converter is not necessarily applicable to the buck converter used in wireless receiver systems. The extra phase lagging attributed to the RHP zero of the buck converter in the wireless power receiver system may exhaust the phase margin and hence lead to instability.

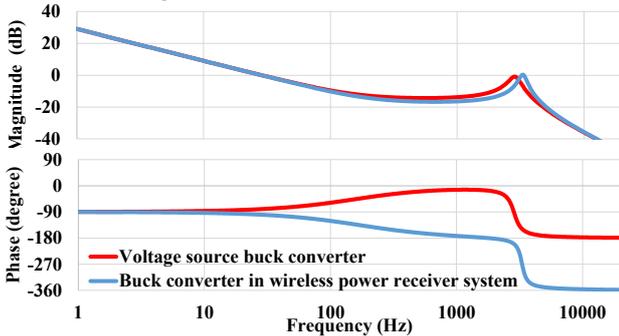

Fig. 8 Bode plots of the closed-loop system of voltage source buck converter and the buck converter in the wireless power receiver system using the same PI compensator.

Fig. 8 show the closed-loop Bode plots of the voltage source buck converter and those of the buck converter in the wireless power receiver system, of which the parameters of the PI compensator are $k_p$=0.1 and $k_i$=10 in both cases. The Bode magnitude plots of both converters are similar. The cut-off frequency is located at around 4771 Hz. However, their Bode phase plots vary significantly. The phase margin of the voltage source buck converter is around 11°, indicating a stable output voltage. However, the phase margin of the buck converter in the wireless power receiver system is −166°, suggesting an unstable output voltage.

*D. Dual-Loop Compensator Design*

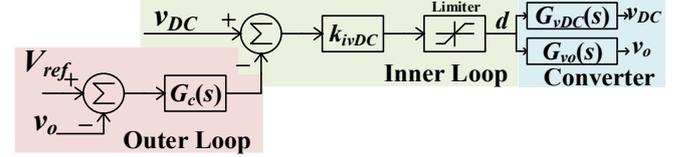

Fig. 9 Block diagram of the dual-loop compensator.

Fig. 9 shows the block diagram of the dual-loop compensator that is utilized to resolve the abovementioned stability problem. The inner loop is introduced to regulate the DC-link voltage, while the outer loop is used to regulate the output voltage. With the introduction of the inner loop, the phase-lagging effect due to the DC-link capacitor can be alleviated and the closed-loop bandwidth for the outer loop can be increased.

The loop gain of the inner loop $LG_i(s)$ can be written as

$$LG_i(s) = -k_{ivDC}G_{vDC}(s) \quad (26)$$

The gain of the inner loop $k_{ivDC}$ is designed based on the $G_{vDC}(s)$ such that the crossover frequency of the inner loop is located at around 1/10 of the switching frequency, i.e. $0.1f$. Correspondingly, $k_{iDC}$ is designed as

$$k_{ivDC} = \left|\frac{1}{G_{vDC}(\mathrm{i}0.2\pi f)}\right| \quad (27)$$

After designing the inner loop, the PI compensator $G_c(s)$ of the outer loop can be designed. The loop gain of the outer loop $LG_o(s)$ is derived as

$$LG_o(s) = \frac{k_p\left(s + \frac{k_i}{k_p}\right)}{s} \times \frac{k_{ivDC}G_{vo}(s)}{k_{ivDC}G_{vDC}(s) - 1} \quad (28)$$

To ensure stability, the roots of $LG_o(s) + 1 = 0$ are located on the left-half plane. By solving this condition, $k_p$ is selected as

$$0 < k_p < \frac{D(CR^2 + L)}{C_{DC}R^2} \quad (29)$$

The zero of the compensators $k_i/k_p$ is designed to locate at 1/20 of the crossover frequency of the outer loop, i.e. $0.005f$. As a result, $k_i$ is designed as

$$k_i = 0.01\pi f k_p \quad (30)$$

By substituting the parameters of the converter, the parameters of the dual-loop compensator are obtained: $k_{ivDC}$=2.3, $k_p$=0.5, and $k_i$=3142. Fig. 10 shows the Bode plots of inner and outer loop gain, respectively. The crossover frequency of the inner loop gain is 20 kHz, and the phase margin is 53°. The crossover frequency of the outer loop gain is 217 Hz, the phase margin is 50°, and the gain margin is 2.49 dB. In addition, the outer loop gain shows a gain of 41.8 dB at 1 Hz. These results suggest that both the inner loop and outer loop are stable with negligible DC regulation error.

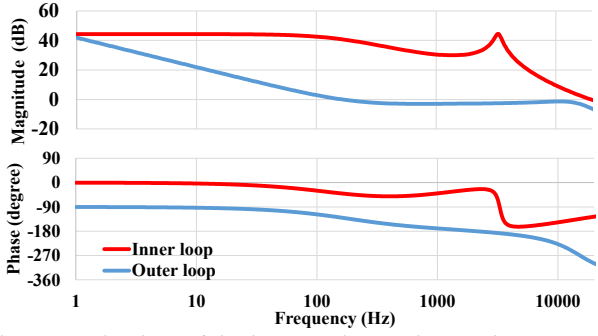

Fig. 10 Bode plots of the inner and outer loop gain.

## IV. EXPERIMENTAL VERIFICATION

TABLE I. LIST OF COMPONENTS

| Component | Value / Part Number |
|---|---|
| $L_s$ | 164 $\mu$H ($d$=29 cm, air core) |
| $C_s$ | 3.86 nF |
| $C_{DC}$ | 30 $\mu$F |
| $L$ | 77 $\mu$H |
| $C_o$ | 40 $\mu$F |
| $R$ | 7 $\Omega$ |
| Gate Driver | ADuM3223 |
| MOSFET | IRLU3714ZPBF |
| Diodes ($D_1$-$D_4$) | IRFB7545 |
| MCU | LAUNCHXL-F28379D |

TABLE II. PARAMETERS OF THE PROTOTYPE

| Parameters | Values |
|---|---|
| Duty ratio $D$ | 50% |
| Switching frequency $f$ | 200 kHz |
| The amplitude of AC input $I_{Ls}$ | 1 A |

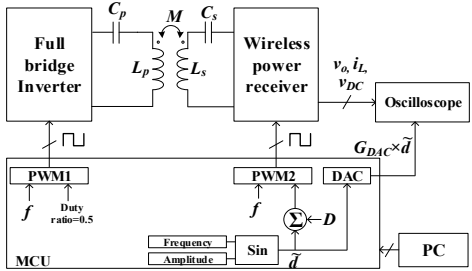

(a) Block diagram of the experimental setup

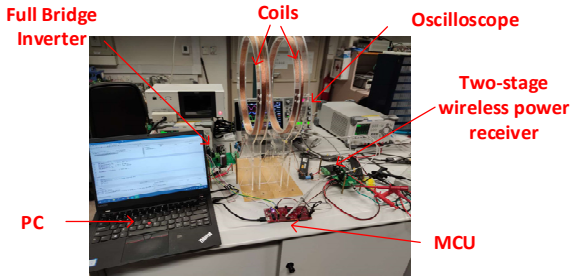

(b) Photograph of the experimental setup

Fig. 11. (a) Schematic diagram and (b) photograph of the prototype and the experimental setup.

To verify the accuracy of the small-signal model, and the existence of the RHP zeros and their effects on the control performance, a prototype of the WPT system is constructed. The components and parameters of the prototype are shown in Table I and II, respectively. Fig. 11 show the experimental setup, where a DSXO3024T oscilloscope, N2873 passive voltage probe, N2790A differential voltage probe, and 1147B current probe are used for the measurement. A microcontroller unit (MCU) is utilized to implement the perturbation via superposing a sinusoidal signal on the nominal duty ratio. Additionally, the perturbation sinusoidal signal is amplified ($\times G_{DAC}$) and converted into an analog signal for phase reference via a digital-to-analog converter (DAC). A personal computer (PC) is connected to the MCU as the master computer, which is used to sweep the frequency and adjust the amplitude of the perturbation sinusoidal signal. The perturbation signal and small-signal responses are observed using the oscilloscope. By measuring the amplitude of the small-signal responses as well as the time-delay between the perturbation and small-signal responses, the Bode plots are obtained.

*A. Time-Domain Steady-State Response*

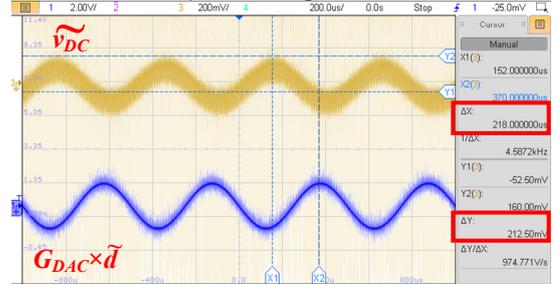

(a) Waveforms of $\widetilde{v_{DC}}$ and $\tilde{d}$

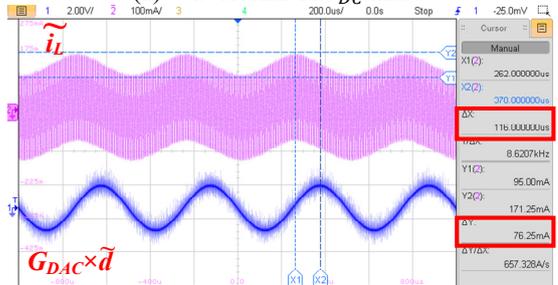

(b) Waveforms of $\tilde{i_L}$ and $\tilde{d}$

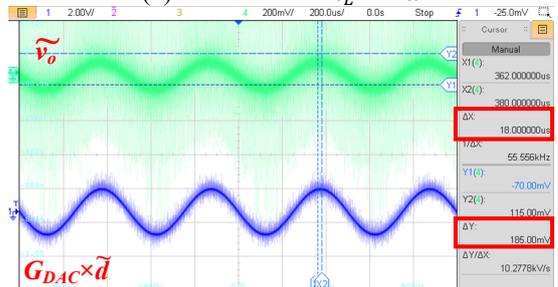

(c) Waveforms of $\widetilde{v_o}$ and $\tilde{d}$

Fig. 12. Waveforms of small-signal AC variations $\widetilde{v_{DC}}$, $\tilde{i_L}$, and $\widetilde{v_o}$ in response to the perturbation $\tilde{d}$.

The measurement of the small-signal response at 2 kHz is shown in Fig. 12 for illustrative purpose. The time waveforms of the output small-signal AC variations of $\widetilde{v_{DC}}$, $\widetilde{i_L}$, and $\widetilde{v_o}$ corresponding to a 2 kHz sinusoidal perturbation $\tilde{d}$ are measured. In this case, the perturbation $\tilde{d}$ is

$$\tilde{d} = 0.005 \times \sin(2\pi \times 2000 \times t) \quad (31)$$

and the gain of the DAC

$$G_{DAC} = 300 \quad (32)$$

The peak-to-peak value of the perturbation, as shown in Fig. 12, is 3 V. Fig. 12(a) shows the waveform of $\widetilde{v_{DC}}$ and $\tilde{d}$. The peak-to-peak value of $\widetilde{v_{DC}}$ is 212.50 mV and the lead time of $\widetilde{v_{DC}}$ is 218 $\mu$s. The measurement results implied a small-signal gain of 26.5 dB and a phase shift of 157°.

Fig. 12(b) shows the time waveforms of $\widetilde{i_{DC}}$ and $\tilde{d}$. The peak-to-peak value of $\widetilde{i_L}$ is 76.25 mA and the lead time of $\widetilde{i_L}$ is 116 $\mu$s. The measurement results implied a small-signal gain of 17.6 dB and a phase shift of 84°.

Fig. 12(c) shows the waveform of $\widetilde{v_o}$ and $\tilde{d}$. The peak-to-peak value of $\widetilde{v_o}$ is 185.00 mV and the lead time of $\widetilde{v_o}$ is 18 $\mu$s. The measurement results implied a small-signal gain of 25.3 dB and a phase shift of 13°.

In summary, the time-domain waveforms given in Fig. 12 verified the feasibility of the measurement of the small-signal gain and phase response at 2 kHz. By adjusting the frequency of perturbation and repeating this process, the Bode plots of small-signal model can be obtained.

### B. Frequency-Domain Response

According to our analysis, most of the poles and zeros of the system are located within the range of 100 Hz to 5 kHz. Consequently, the measured frequency range (from 10 Hz to 10 kHz) is sufficient for verification of the Bode plots. Fig. 11(a), (b) and (c) show the experimental and theoretical Bode plots of $G_{v_{DC}}(s)$, $G_{i_L}(s)$, and $G_{v_o}(s)$, respectively. The measured Bode plots are generally close to the theoretical ones.

Fig. 13(a) shows the experimental and theoretical Bode plots of $G_{v_{DC}}(s)$. It shows a slight phase drop of around 36° between 10 Hz and 200 Hz as well as a significant phase drop of around 180° between 2 kHz to 5 kHz. This suggests the presence of an LHP real pole located at approximately 200 Hz and a pair of conjugate LHP poles located between 2 kHz and 5 kHz. Besides, there are phase increments around 1 kHz and 10 kHz which implies that there are two LHP zeros.

Fig. 13(b) shows the experimental and theoretical Bode plots of $G_{i_L}(s)$. It shows a slight phase drop of around 100 Hz and a significant phase drop of around 180° at around 2 kHz. This suggests the presence of an LHP real pole located at approximately 200 Hz and a pair of conjugate LHP poles located between 2 kHz and 5 kHz. The RHP zero (at 189 Hz) and the LHP zero (at 568 Hz) are very close, thereby flattening the frequency region of the Bode phase plot.

Fig. 13(c) shows the experimental and theoretical Bode plots of $G_{v_o}(s)$. It shows two significant phase drops of 180° at around 100 Hz and 2 kHz. The phase drop at around 2 kHz is attributed to the conjugate poles while the phase drop at around 100 Hz is contributed by a real pole and a RHP zero. As a consequence, the RHP zero is located at around 100 Hz.

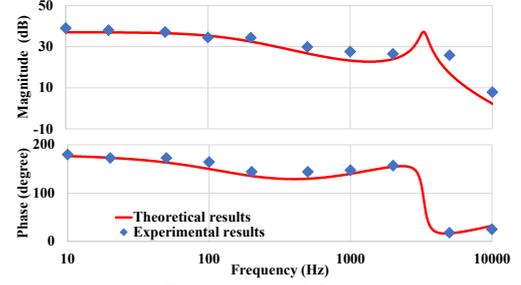

(a) Bode plots of $G_{v_{DC}}(s)$

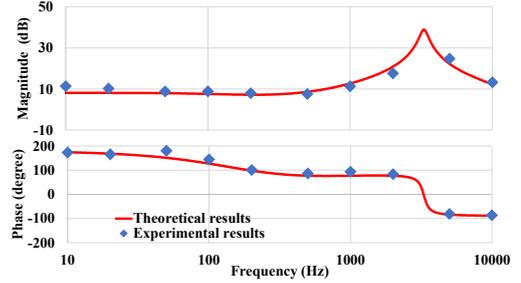

(b) Bode plots of $G_{i_L}(s)$

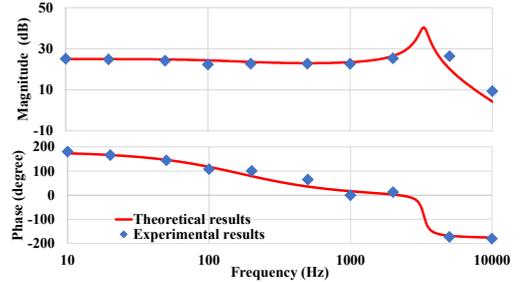

(c) Bode plots of $G_{v_o}(s)$

Fig.13. Experimental and theoretical Bode plots of $G_{v_{DC}}(s)$, $G_{i_L}(s)$, and $G_{v_o}(s)$

### C. Time-Domain Open-Loop Dynamic Response

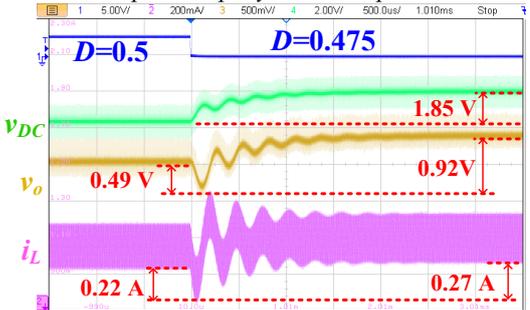

Fig.14. Dynamic response of the buck converter in response to the step change of duty ratio.

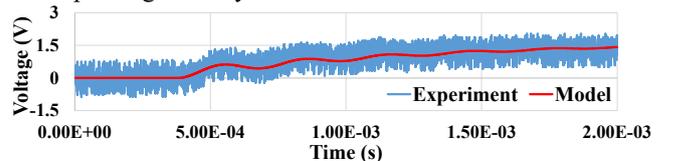

(a) Step responses of $v_{DC}$

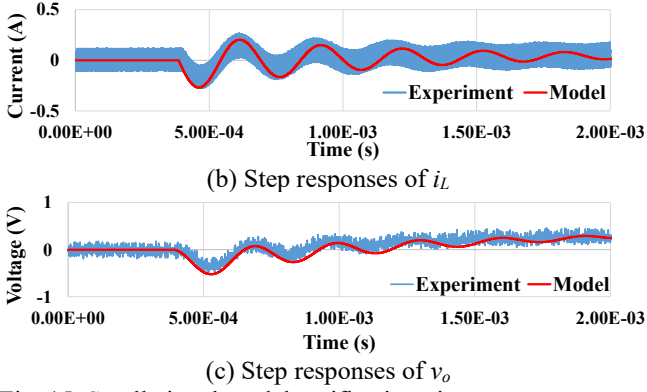

(b) Step responses of $i_L$

(c) Step responses of $v_o$

Fig. 15. Small-signal model verification via step responses.

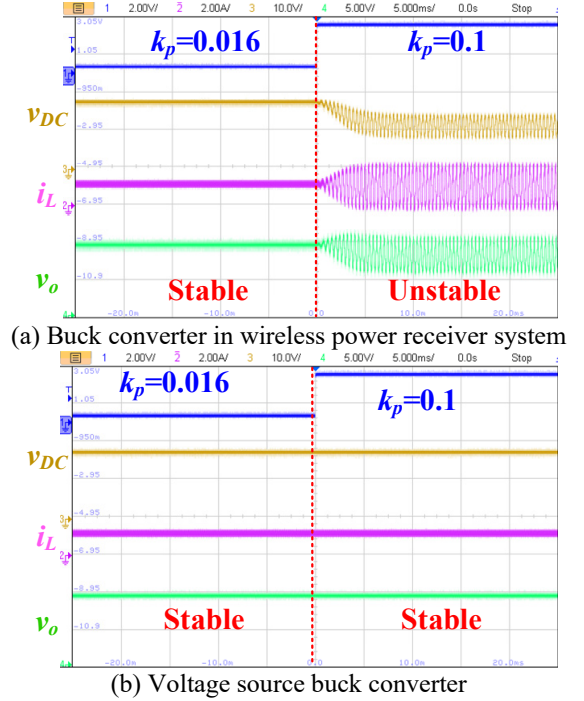

(a) Buck converter in wireless power receiver system

(b) Voltage source buck converter

Fig. 16. Closed-loop dynamic responses of the (a) buck converter in wireless power receiver and (b) voltage source buck converter for a step change in $k_p$.

Fig. 14 shows the open-loop dynamic response of the buck converter in response to the step change, where the duty ratio is stepped down from 0.5 to 0.475. After the step change of duty ratio, $v_{DC}$ increases by 1.85 V. Meanwhile, $v_o$ initially drops by 0.49 V, before increasing by 0.92 V to reach the new steady-state value. Similarly, $i_L$ initially decreases by 0.22 A, then rises by 0.27 A to reach its new steady-state value. The monotonical dynamics of $v_{DC}$ suggests the absence of RHP zero in $G_{v_{DC}}(s)$, whereas the non-monotonical dynamics of $i_L$ and $v_o$ verify the existence of RHP zeros in $G_{i_L}(s)$ and $G_{v_o}(s)$ and how they are affecting the dynamic response. Moreover, Figs. 15 (a), (b), and (c) respectively show the dynamic response waveforms of $v_{DC}$, $i_L$, and $v_o$ with respect to a duty cycle step change from 0.5 to 0.475. The experimental waveforms (in blue) and the waveforms of the model (in red) are quite close, validating the accuracy of the small-signal model.

*D. Time-Domain Closed-Loop Dynamic Responses*

Fig. 16 gives an experimental comparison of the time-domain closed-loop dynamic responses of a voltage source buck converter and that of the same buck converter being used in the wireless power receiver system. In this experiment, the proportional gain of the PI compensator is step-changed from $k_p$=0.016 to $k_p$=0.1. At $k_p$=0.016, both converters are stable. However, after the step change to $k_p$=0.1, the waveforms of the buck converter in the wireless power receiver system become unstable, while the voltage source buck converter remains stable. The result also reflects that with the RHP zero, the buck converter in the wireless power receiver system has a relatively narrow stability region. As a consequence, it is critical to account the RHP zero in the compensator design of this system.

A comparative study of different compensator design methods [17], [18] for the buck converter of the wireless power receiver is shown in Table III. Unlike the frequency-domain small-signal model [17], the model proposed in this work exposes the presence of the non-minimum-phase dynamics such that instability due to the RHP zero can be avoided via proper design. As compared to the time-domain large-signal model [18], this work features model-based compensator parameter design and simpler stability assessment, which facilitates a better understanding of closed-loop stability.

TABLE III.
COMPARISON WITH DIFFERENT COMPENSATOR DESIGN METHODS

|  | **This work** | [17] | [18] |
|---|---|---|---|
| Model | **Frequency-domain small-signal model** | Frequency-domain small-signal model | Time-domain large-signal model |
| Compensator | **Frequency-domain PI compensator** | Frequency-domain PI compensator | Time-domain model predictive controller |
| Modulation | **PWM** | PWM | PWM |
| Model complexity | **One 3rd order model** | One 2nd order model | Two 3rd order models |
| RHP zero | **Yes** | No | N/A |
| Model accuracy | **High** | Poor | High |
| Compensator complexity | **1st order** | 1st order | 2nd order or higher |
| Compensator parameter | **Model-based design** | Model-based design | N/A |
| Stability assessment | **Simple** | Simple | Difficult |
| Stability boundary | **Accurate** | Inaccurate | N/A |

V. CONCLUSIONS

In this paper, a wireless power receiver system that is based on the full-bridge rectifier cum buck converter configuration, which takes in a current source input from the wireless receiver coil, is modeled and analysed. It is discovered in this work that the current source nature of the input, which is attributed to the use of series compensation and a finite DC-link capacitance, inherently introduces a right-half-plane (RHP) zero to both the small-signal inductor-current-to-duty-ratio $G_{i_L}(s) = \widetilde{i_L}/\tilde{d}$ and output-voltage-to-duty-ratio $G_{v_o}(s) = \widetilde{v_o}/\tilde{d}$ transfer functions of the buck converter. As shown in the time waveforms obtained via simulation and experiment work, the RHP zeros causes the dynamic response of the inductor current and output voltage to follow a non-monotonic trajectory in the event of a change of the duty cycle. The presence of the RHP zeros have been verified in the frequency domain by experiment and simulation, of which the experimental Bode plots are in close agreement with the analytical ones. This work also provides explanation of the non-minimum-phase dynamic performance of the buck converter in the wireless receiver system and offers guideline on how such compensators can be designed.